# Evidence for the role of the magnon energy relaxation length in the Spin Seebeck Effect

Arati Prakash[1], Benedetta Flebus[2], Jack Brangham[1], Fengyuan Yang[1], Yaroslav Tserkovnyak[3], Joseph P. Heremans[4,1,5]

[1] Department of Physics, The Ohio State University, Columbus, Ohio 43210, USA

[2] Institute for Theoretical Physics and Center for Extreme Matter and Emergent Phenomena, Utrecht University, Leuvenlaan 4, 3584 CE Utrecht, The Netherlands

[3] Department of Physics and Astronomy, University of California, Los Angeles, California 90095, USA

[4] Department of Mechanical Engineering, The Ohio State University, Columbus, Ohio 43210, USA

[5] Department of Materials Science and Engineering, The Ohio State University, Columbus, Ohio 43210, USA

**Abstract**

Temperature-dependent spin-Seebeck effect data on Pt|YIG ($Y_3Fe_5O_{12}$)|GGG ($Gd_3Ga_5O_{12}$) are reported for YIG films of various thicknesses. The effect is reported as a spin-Seebeck resistivity (SSR), the inverse spin-Hall field divided by the heat flux, to circumvent uncertainties about temperature gradients inside the films. The SSR is a non-monotonic function of YIG thickness. A diffusive model for magnon transport demonstrates how these data give evidence for the existence of two distinct length scales in thermal spin transport, a spin diffusion length and a magnon energy relaxation length.

Since the discovery of the (longitudinal) spin-Seebeck effect (SSE)[1], much work has been done to identify the length scales involved in the phenomenon.[2] Using nonlocal detection, it has been shown that relaxation of thermal magnons in YIG is governed by a spin diffusion length.[3] The latter is reported to be around 10 μm, and, in some studies, increases to up to 70 μm at low temperatures.[4,5]. This has led to a consensus that the micron-scale dependence of SSE observed in planar geometries corresponds to the generation and accumulation of the nonequilibrium magnon-density gradients in the bulk. These experiments have been modeled theoretically in terms of magnon spin transport only, while assuming that the magnon-phonon processes leading to the relaxation of the magnon energy occur over very short length scales; hence, their effects can be disregarded.[3,5] Nevertheless, it is clear that magnon energy relaxation mechanisms by the phononic environment must be invoked generally for a complete understanding of thermal spin transport, and particularly for the physics underlying the SSE. Indeed, while heaters and thermometers couple to phonons, these must in turn couple to magnons in order to give rise to the SSE in a magnon-based system like YIG. These relaxation processes can be parameterized by the length over which magnon-to-phonon thermalization occurs, and the latter is expected to be a much smaller length scale than magnon spin-diffusion lengths; a theoretical argument can be found in Ref. [6]. Early work[7] defines an energy relaxation length similar to the one invoked here, but without quantifying it. Additionally, phonon-magnon drag has been put into evidence in previous SSE experiments[8,9], which, again, points to the importance of interactions between magnons and phonons. To the best of our knowledge, no explicit evidence for the effect of this length scale on SSE measurements has been reported to date.

Previous articles on thin films using various growth techniques[10,11,12] have shown the SSE signal to increase with increasing YIG film thickness. In this study, we grow a series of Pt|YIG|GGG heterostructures, with YIG thickness varying from 10 nm to 1 μm, using the same growth technique for all films. We measure the temperature-dependent spin-Seebeck effect on these structures and of bulk single-crystal Pt|YIG. The spin-Seebeck signal increases for film thicknesses from 10 to 250 nm and again for the bulk YIG film, but in between, the signal reaches a local maximum at a thickness ~ 250 nm – a detailed comparison with previous studies is below.

We explain the non-monotonic behavior in terms of the energy-equilibration dynamics of magnons to phonons in the YIG. We implement a diffusive model for spin and heat transport in order to parametrize magnon-phonon thermal relaxation,[6,7] which allows us to interpret our observations. Using typical material parameters for YIG and values for interfacial thermal conductances of the Pt|YIG|GGG heterostructure from Ref. [13], we calculate thermally driven spin current as a function of the YIG film thickness. In our picture, a local maximum is governed by the magnon energy relaxation length, and the spin current eventually saturates at thicknesses beyond the measured magnon spin-diffusion length. Thus, the non-monotonicity is evidence that there are two mechanisms at play, occurring on two different length scales, which define the magnonic spin-Seebeck effect.

A series of YIG films with varied thickness (10 nm, 40 nm, 100 nm, 250 nm, 500 nm, 1 μm) were grown epitaxially on single crystal GGG <100> substrates by ultrahigh-vacuum off-axis sputtering. Then, Pt layers of thickness 6 nm were deposited at room temperature on the YIG films, also by off-axis sputtering. A bulk, 500 μm single-crystal YIG <100> substrate was obtained from Princeton Scientific, so that YIG crystal orientation was controlled for all samples. Pt was deposited on this bulk crystal in the manner described for GGG. We verified that the quality of all film thicknesses used for SSE measurements is uniform. This was done by vibrating sample magnetometer measurements of the saturation magnetization and by measurements of the ferromagnetic resonance linewidth. The data on the films up to 250 nm thickness are in Refs. [14] and [15]. The supplement reports additional measurements of the saturation magnetization, extending the results to YIG films of 1 μm thickness.

Spin-Seebeck thermopower measurements were conducted at low temperatures from 2 to 300 K on a Quantum Design Physical Property Measurement System (PPMS).[16] Here, the electrical measurements are quite accurate. However, the measurements of the imposed temperature profile are not as accurate because this quantity must be estimated across the layer of YIG that supports magnon transport. Practical measurements either obtain the temperature difference $\Delta T$ across the full Pt|YIG|GGG stack, and proportion this somehow across the layers, or estimate a value for the temperature gradient $\nabla T$ from a heat flux measurement combined with a known value of thermal conductivity. Crucially, the estimates of $\nabla T$ require knowledge of its variation across the Pt|YIG|GGG stack, which is inaccessible experimentally given the presence of interfacial thermal resistances (Pt|YIG and YIG|GGG), the differences between the thermal conductivities of the different layers, and that the thermal conductivity of the YIG film is not known accurately since it depends on thickness. We show in detail in the supplement how these methods yield inconsistent conclusions. In particular, the temperature dependence of the experimental results when represented as a spin-Seebeck coefficient (SSC) in thermopower units [μV/K] reflects, in essence, the temperature dependence of the thermal conductivity of bulk YIG or GGG, which is dominated by phonons above 10 K.[17] Further, recent work[18] systematically shows the lack of repeatability and, thus, reliability of this representation of the magnitude of the SSE.

In a proper adiabatic sample mount, the heat flux $j_Q$ (in units of W/m$^2$) is unidirectional, flows entirely into the sample, and is measured reliably from the knowledge of the sample cross-section and the amount of electrical power dissipated by a resistive heater. Cryostat calibrations show the heat losses to be at most 15 mW/K at 300 K, slightly less than two orders of magnitude lower than the thermal conductance of the sample, so that $j_Q$ is a well-defined experimental quantity by which we can parameterize the spin-Seebeck response.

Circumventing the need to estimate $\nabla T$, we report our data in terms of a spin-Seebeck resistivity (SSR), $R_{SSE} \equiv E_{Pt}/j_Q$ (in units of nm/A), as a function of temperature in Figure 1. From these data, we derive the thickness dependence of SSR over a broad temperature range, spanning 20 to 300 K, in Figure 2. The data show two features. The first is the non-monotonic behavior of the thickness dependence of the SSR described above: the signal shows first an increase with increasing thickness up to 250 nm, but is followed by a clearly resolved decrease, leading to a minimum at or slightly above 1 μm. Second, while the data on bulk YIG clearly are dependent on temperature, those below 1 μm have only a weak dependence. We point out that this feature is seen best when the data are plotted as SSR (per unit heat flux), and less so in the SSC (μV/K), which can be seen in the supplement.

These results are discussed in the context of the existing literature. Ref. [10] studies thickness dependence of the SSE on two different sample sets: the results suggest a saturation of the SSE signal at 200 nm for the first set, and around 10 μm for the second. Film thicknesses in the pertinent interval (between 200 nm and several microns) are not studied. The data from Ref. [10] are thus broadly consistent with the observations of Fig. 2. The non-monotonicity appears to be absent in Ref. [11] in the relevant thickness range. However, there are rather few data points in the relevant thickness range, and the SSE signals are an order of magnitude lower than those presented here (in units of nV/K in the supplement) so that the amplitude falls quite close to the apparent resolution in Ref. [11]. These factors would make it difficult to resolve the local maximum reported in Fig. 2. Finally, in Ref [12], which uses a different measurement method, the thickness dependence of the SSE voltage shows one outlying data point at 200 nm, which corresponds rather well to the observation in Fig. 2. This point also deviates from the model, which is based only on a magnon diffusion length. In fact, the discussion in Ref [12] concludes that the local SSE may be governed by a length scale different from the magnon diffusion length that is extracted from nonlocal signals. From this comparison to the literature, it appears that Fig. 2 offers insight to such a length scale, by adding data points in the relevant YIG thickness range.

We model the experiment by considering the one-dimensional geometry shown in Fig. 3. The experimentally controlled heat flux defines the phonon temperature gradient $\partial_x T_p < 0$ in YIG. We treat phonon transport as diffusive and we assume that the phonon temperature in the YIG, $T_p$, is not perturbed significantly by the magnons. The temperature drops at the left interface, $\delta T_L = T_{p,L} - T_p(0)$, and at the right, $\delta T_R = T_p(d) - T_{p,R}$, with $T_{p,L(R)}$ being the phonon temperature in Pt (GGG), can then be determined by solving the phonon heat-diffusion equation[13]:

$$\delta T_{L(R)} = -\ell_{L(R)} \, \partial_x T_p . \qquad (1)$$

Here, we have introduced the Kapitza length of the Pt(GGG)|YIG interface as $\ell_{L(R)} = \kappa_p/\mathrm{K}_{p,L(R)}$, where $\kappa_p$ is YIG phonon thermal conductivity and $\mathrm{K}_{p,L(R)}$ the phonon interfacial thermal conductance for the Pt(GGG)|YIG interface.

In YIG, scattering processes among thermal magnons, which are governed by exchange interactions, occur on a much shorter timescale than magnon lifetime. Magnons are then described by a magnon temperature $T_m(x)$ and an effective magnon chemical potential $\mu(x)$[5,19] that parametrize the non-equilibrium distribution induced by the thermal bias. Treating transport semi-classically, the magnon spin and heat continuity equations read as[6]

$$\partial_t n + \partial_x j = -g_{n\mu}\,\mu - g_{nT}\left(T_m - T_p\right), \tag{2a}$$

$$\partial_t u + \partial_x q = -g_{uT}\left(T_m - T_p\right) - g_{u\mu}\,\mu, \tag{2b}$$

where $n$ is the density of the thermal magnons and $u$ the energy density carried by them. The $g_{n\mu}$ and $g_{u\mu}$ coefficients account for the relaxation by phononic environment of magnon spin and temperature, respectively. The cross-terms $g_{nT}$ and $g_{uT}$ describe the generation or decay of spin accumulation by heating or cooling of magnons and vice versa, and are related by the Onsager-Kelvin relation, i.e., $T_m\,g_{nT} = g_{u\mu}$. In the linear response regime and neglecting magnon-phonon drag contributions $j, q \propto \partial_x T_p$ (restoring phonon drag would simply rescale the bulk bias terms proportional to $\partial_x T_p$), the spin, $j$, and heat, $q$, currents carried by magnons can be written as

$$j = -\sigma \partial_x \mu - \zeta \partial_x T_m, \tag{3a}$$

$$q = -\kappa \partial_x T_m - \varrho \partial_x \mu, \tag{3b}$$

where $\sigma, \kappa, \zeta$, and $\varrho = T_m\,\zeta$ are the bulk spin and heat conductivities and the intrinsic spin Seebeck and Peltier coefficients, respectively.

Equations (2a) and (2b) must be determined consistently with the boundary conditions for spin and heat transport at the Pt|YIG and YIG|GGG interfaces. For Pt|YIG, at $x = 0$, the latter reads as

$$j = -G_L\,\mu + S_L\,\delta T_L + S_L(T_p - T_m), \tag{4a}$$

$$q = K_L\,\delta T_L + K_L\,(T_p - T_m) - \Pi_L\,\mu, \tag{4b}$$

where $G_L$, $K_L$, $S_L$, and $\Pi_L = T_m S_L$ are the interfacial magnon spin and thermal conductances and spin-Seebeck and Peltier coefficients, respectively. Note that the spin current (4a) injected at the Pt|YIG interface is directly proportional to the measurable inverse spin-Hall voltage.[20]

At the YIG|GGG interface, the spin flow is blocked as there are no spin carriers in the GGG substrate. Nevertheless, heat still can be transmitted via inelastic spin-preserving scattering processes between magnons and phonons. The corresponding boundary conditions at $x = d$ can be written as:

$$j = 0, \tag{5a}$$

$$q = K_R \, \delta T_R - K_R \, (T_p - T_m). \tag{5b}$$

Here, $K_R$ is the interfacial magnon heat conductance, which accounts for the processes leading to energy exchange between magnons and phonons at the YIG|GGG interface. The interfacial magnon spin conductance $G_R$, the spin Seebeck $S_R$, and, consequently, the spin-Peltier coefficient $\Pi_R = T_m S_R$ vanish.

Next we introduce the magnon energy relaxation length, $\lambda_T = \sqrt{\kappa/g_{uT}}$, and the magnon spin-diffusion length, $\lambda_s = \sqrt{\sigma/g_{n\mu}}$, which parametrize the relaxation of the magnon temperature to the phonon temperature and the relaxation of the magnon chemical potential to its equilibrium (vanishing) value, respectively. Since the spin-preserving relaxation of magnon distribution towards the phonon temperature does not rely on relativistic spin-orbit interactions, the associated length scale $\lambda_T$ can be taken to be much shorter than the spin diffusion length $\lambda_s$.[5,6,21]

Based on the latter assumption, we identify two distinct transport regimes as functions of the thickness $d$ of the YIG. When the thickness is comparable or larger than the spin-diffusion length, i.e., $d \geq \lambda_s \gg \lambda_T$, we assume the transport is dominated by the mechanisms leading to the relaxation of the magnon chemical potential, while setting $T_p = T_m$ throughout the YIG. [See supplementary material]. As shown in Figure 4, the corresponding contribution to the injected current at the Pt|YIG interface grows monotonically as function of the YIG thickness, ultimately to saturate for large thicknesses. The underlying physical picture is clear. The temperature bias induces a chemical potential imbalance at the left interface. The corresponding magnon density increases with increasing sample thickness. However, when $d \gg \lambda_s$, magnons will decay – via non-spin-preserving interactions with the lattice – before reaching the left end of the sample, and the SSE signal will not increase further.

In the opposite regime, when $d \sim \lambda_T \ll \lambda_S$, we instead disregard any chemical-potential imbalance (i.e, by setting $\mu = 0$). In this case, the relaxation of the magnon temperature to the phonon one emerges as a driving mechanism for transport. In order to analyze the corresponding contribution to the injected spin current, we treat the magnon heat Kapitza length $\ell^*_{T,R} = \kappa/K_R$ at the YIG|GGG interface as a free parameter, as there is no current estimate for interfacial thermal conductance $K_R$. Estimates for the other length scales can be found in the supplementary material. Figure 5 shows that we reproduce a non-monotonic behavior of the injected spin current for $\ell^*_{T,R} \ll \ell^*_{T,L}, \lambda_T$. To understand this result, recall that the magnon heat Kapitza length $\ell^*_{T,R}$ parametrizes the strength of spin-preserving inelastic interactions between magnons in YIG and phonons at GGG at the right interface; in other words, a short $\ell^*_{T,R}$ corresponds to an efficient thermalization process. As the phonon temperature in the GGG substrate $T_{p,L(R)}$ is lower than that in the YIG, i.e., $\delta T_R > 0$, thermalization mechanisms between magnons in YIG and phonons in GGG lower the magnon temperature $T_m$ near the YIG|GGG interface over a shorter timescale than the one governing the thermalization of magnons and phonons in YIG. Hence, the right interface acts as a source of "cold" magnons, which contribute positively to the temperature imbalance in YIG, $T_p - T_m$. However, when the sample thickness is larger than the temperature relaxation length, $\lambda_T$, these magnons thermalize with phonons before reaching the left interface. Thus, for $d > \lambda_T$, the YIG|GGG interface no longer acts as an effective source for the magnon-phonon temperature imbalance in YIG, leading to the non-monotonicity observed in Figure 5. We suggest that this theoretical feature is related to our experimental observations and that it implies that the thickness at which the peak in SSR versus YIG thickness emerges is in direct correspondence with the magnon energy relaxation length in YIG, leading to an experimental estimate of the latter as $\lambda_T \sim 250$ nm.

Relating the experimental peak to the magnon energy relaxation length scale allows us to interpret the weak temperature dependence of the SSR measurements as the little-to-none temperature dependence of the magnon energy relaxation length. While surprising at first, the apparent lack of temperature dependence of $\lambda_T$ shown in Figure 2 parallels that reported for the spin-diffusion length $\lambda_s$ [3]. Additionally, although the microscopic origin of $\lambda_s$ relies on spin-relaxing interactions, whereas the magnon energy relaxation length $\lambda_T$ is in our analysis introduced as a spin-preserving interaction, it may be interesting to compare the weak temperature dependence of $\lambda_T$ to the rather weak temperature dependence of the Gilbert damping parameter.[22,23] Notwithstanding these remarks, the weakness of the temperature dependence of $\lambda_T$ is an issue that should be investigated in the future.

In summary, our data and model suggest that the measured dependence of the SSR on the sample thickness reflects magnon spin- and heat-diffusion processes occurring on separate length scales. We show how a non-monotonic feature in the signal observed at short thicknesses can emerge corresponding to the length scale parametrizing magnon-phonon thermalization mechanisms, and we offer an estimate for the latter, i.e., $\lambda_T \sim 250$ nm. While a diffusive approach to the magnon heat transport qualitatively captures the newly observed feature, here we want to stress that our model represents a concept, rather than a complete theory of transport, at short thicknesses. Future work should address the nonequilibrium mechanisms operating over such length scales beyond the diffusive approximation rigorously.

**Acknowledgement**

This work was supported primarily by the Army Research Office (ARO) MURI W911NF-14-1-0016 and the U.S. Department of Energy (DOE), Office of Science, Basic Energy Sciences, grant No. DE-SC0001304. It is also supported by the Center for Emergent Materials, an NSF MRSEC, grant DMR-1420451. Additional funding is from the European Research Council, and the D-ITP consortium, a program of the Netherlands Organization for Scientific Research (NWO) that is supported by the Dutch Ministry of Education, Culture and Science (OCW).

**Figures**

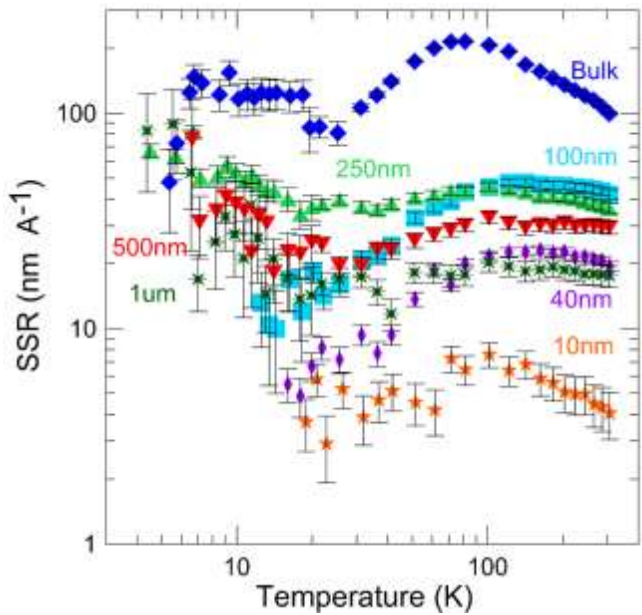

Figure 1. Temperature dependence of the SSR for various YIG film thicknesses and bulk YIG.

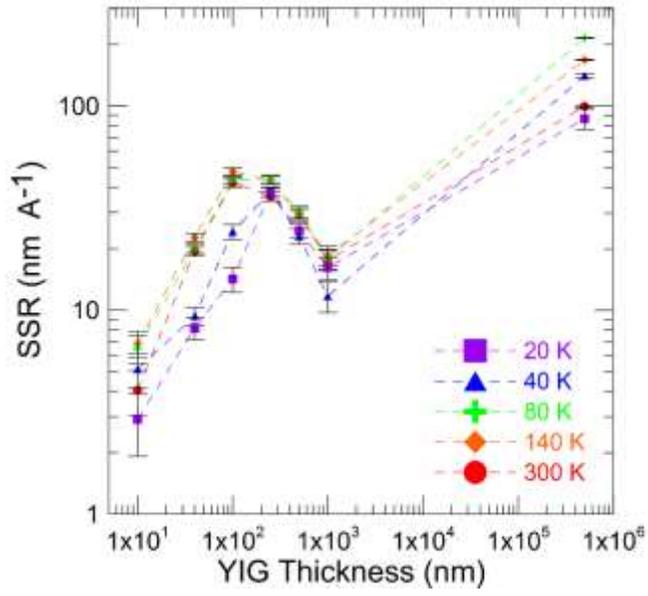

Figure 2. Thickness dependence of the SSR at the various temperatures indicated on the graph.

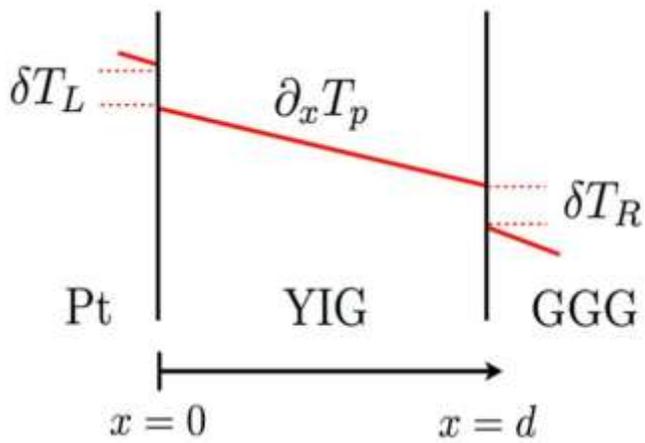

Figure 3. Qualitative illustration of the phonon temperature profile in a Pt|YIG|GGG heterostructure.

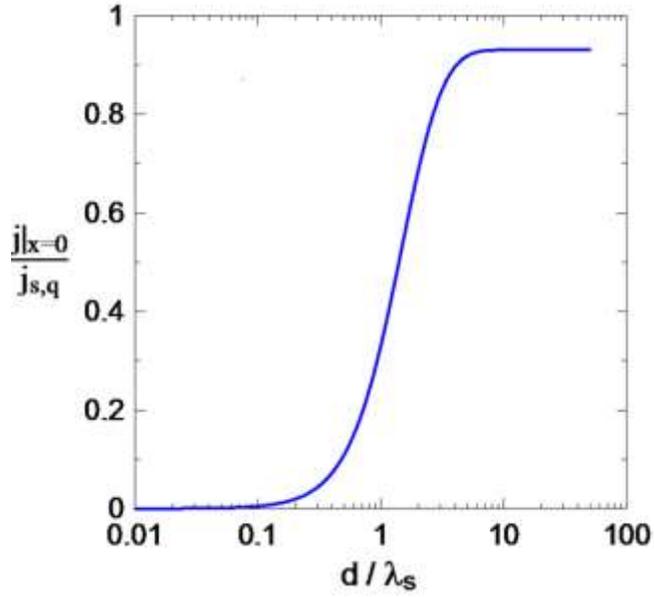

Figure 4. Thickness dependence of the spin current injected at the Pt|YIG interface in the far regime, i.e. where the thickness is represented on the scale of the magnon spin diffusion length $\lambda_S$. Here, spin transport is dominated by the mechanisms leading to relaxation of the magnon chemical potential. The spin current is normalized by the thermal spin current in the bulk $j_{s,q} = -\zeta \partial_x T_p$.

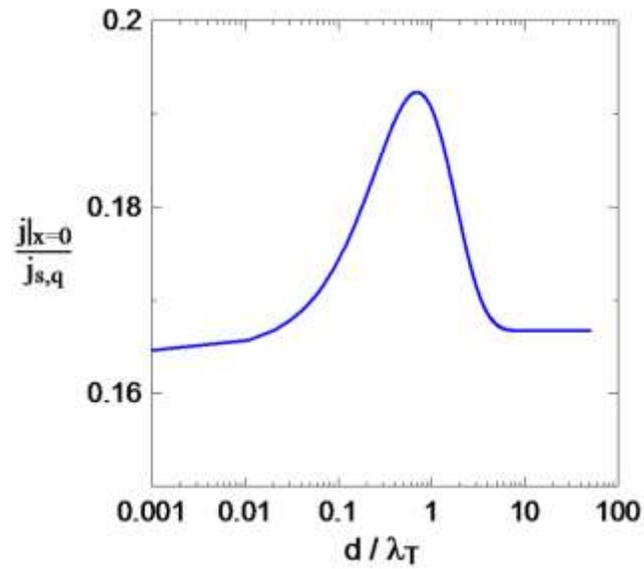

Figure 5. Thickness dependence of the spin current injected at the Pt|YIG interface in the near regime, i.e., where the thickness is represented on the scale of the magnon energy relaxation length $\lambda_T$. Here, the chemical-potential imbalance is disregarded, and the non-monotonicity results from the interplay between $\lambda_T$ and the presence of cold magnons at the YIG/GGG interface. The spin current is normalized by the thermal spin current in the bulk $j_{s,q} = -\zeta \partial_x T_p$.

**Supplementary Material**

This supplement has three sections. First, we present the data of Figures 1 and 2 of the main text in terms of a spin-Seebeck thermopower, to illustrate the limitations of that approach. Second, we present vibrating sample magnetometry (VSM) data to show that the saturation magnetization of all samples is the same, which, together with ferromagnetic resonance (FMR) data in previous publications, establishes that the sample quality remains uniform for all thicknesses studied. Third, we present a detailed discussion of the theoretical model and the analytical derivation of the expressions plotted in Figures 4 and 5 of the main text.

**Quantifying Temperature Gradients**

The spin-Seebeck intensity is defined conventionally as a thermopower in units of [V/K]: either as the voltage across the Pt film (typically on the order of nano- to microvolts) divided by the measured temperature difference $\Delta T$ (K) applied, or as the electric field E (V/m) in the Pt divided by the temperature gradient $\nabla T$ (K/m) across the Pt|YIG|GGG trilayer. As an alternative (but equivalent) way to represent thermopower, some groups report the signal in units of voltage as a linear function of the applied power or temperature difference at a given temperature, with the slope of the line representing the spin Seebeck coefficient [V/K] at that temperature.

In these methods, the temperature drop is measured directly across the trilayer. However, by following this protocol, one includes temperature drops arising from elements of the thermal stack external to the sample (e.g. heaters, heat spreaders) and/or internal to the trilayer at the Pt|YIG and YIG|GGG interfaces, as well as in the bulk GGG substrate, when evaporated Pt thermistors are used for thermometry. Thus, this type of measurement reflects a temperature profile extraneous to the one of interest, i.e. the one in the YIG, whose magnons give rise to the spin-Seebeck effect. The data from the main text represented using this method are shown Fig. S1.

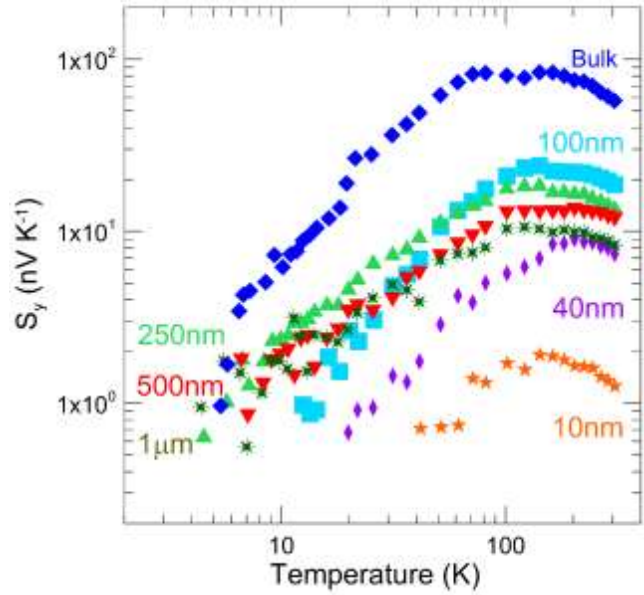

*Figure S1. Temperature dependence of the spin-Seebeck data in units of thermopower using measured values for the temperature difference across the heterostructure.*

One improvement over this method is to not measure the temperature drop at all, but derive $\nabla T$ from the measured heat flux $j_Q$ (in units of W/m$^2$), which is in turn derived from the heater power output $Q$ per unit cross-sectional area of YIG. When the sample is in a proper adiabatic mount in a thermal vacuum (< 10$^{-5}$ Torr), in an environment with gold-plated radiation shields, and contacted only with ultrathin (< 25 μm diameter) thermometer and voltage wires, this heat flux is unidirectional and flows into the sample. A resistive heater dissipates power via a copper heat spreader, and the sample is mounted on a copper heat sink. Using the measured values for bulk thermal conductivity $\kappa_{YIG}$ of YIG at each temperature, the average temperature gradient across the sample can be estimated from Fourier's law $\nabla T = j_Q / \kappa_{YIG}$. The results are shown in Fig. S2.

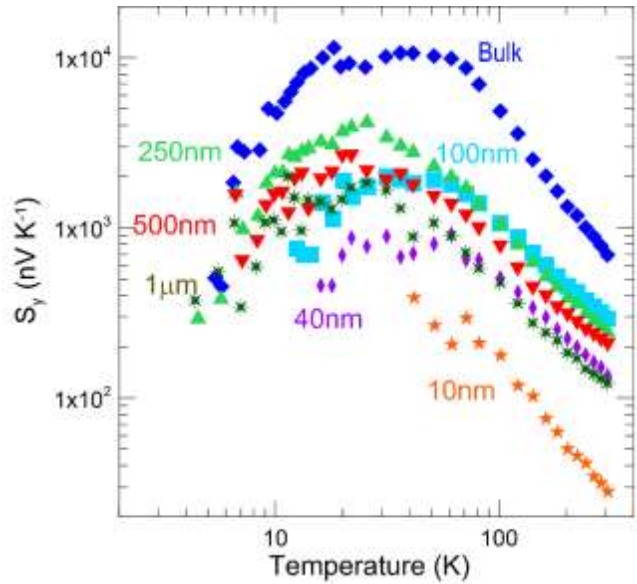

*Figure S2. Temperature dependence of the spin-Seebeck data in units of thermopower using calculated values for the temperature gradient by considering the thermal conductivity of bulk YIG.*

While more reliable than the direct temperature measurement, as it avoids errors arising from thermal contact resistances, this procedure also assumes that the thermal conductivity of every YIG film, from 10 nm to 1 μm, is the same as that of bulk YIG. This assumption is problematic, as it is well known that the thermal conductivity of thin films can differ greatly from bulk values. In the absence of reliable data on the thickness dependence of thermal conductivity of YIG films, this method convolutes the desired measurement of the temperature gradient across the YIG film with a theoretical value of thermal conductivity in bulk YIG. Furthermore, it ignores the effects of Kapitza resistances, which are important to the interpretations of the results. Next, we compare the temperature dependences shown in Fig. S1 and S2 to the thermal conductivity of YIG and GGG, which we measured on bulk samples and reproduce in Fig. S3.

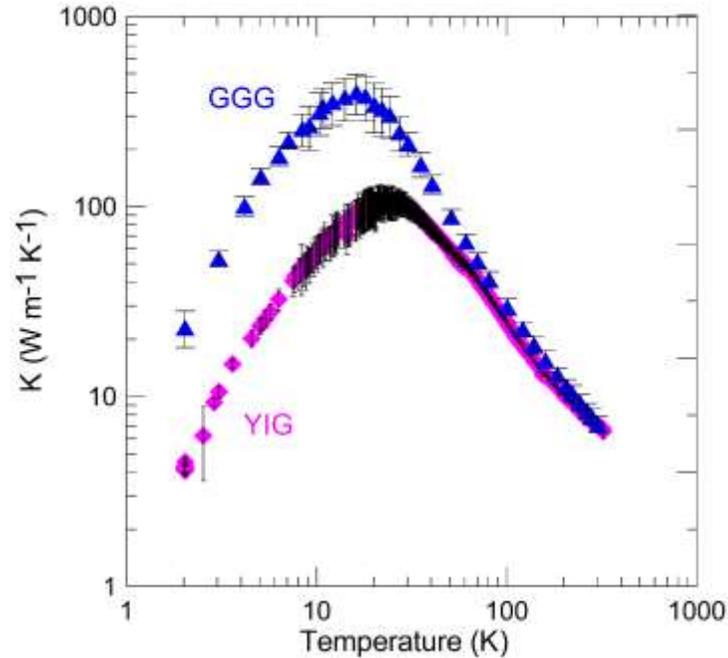

*Figure S3. Temperature dependence of the thermal conductivity of bulk YIG and GGG*

The fact that the temperature dependence of the spin-Seebeck signal reported in Fig. S2 is quite similar to the bulk thermal conductivity of GGG or YIG, which are dominated by phonons, reveals that these methods over-emphasize the role of phonons in spin Seebeck physics. This constitutes an additional argument to represent the spin-Seebeck effect in terms of a spin-Seebeck resistivity, as is done in the main text.

The data from Fig. S2 are represented as a function of YIG thickness in Fig. S4, which can be contrasted with Fig. 2 in the main text. The thickness dependence of the spin-Seebeck resistivity has a much weaker temperature dependence than that in Fig. S4, suggesting that this apparent temperature dependence originates from phonon physics, and has little physical meaning.

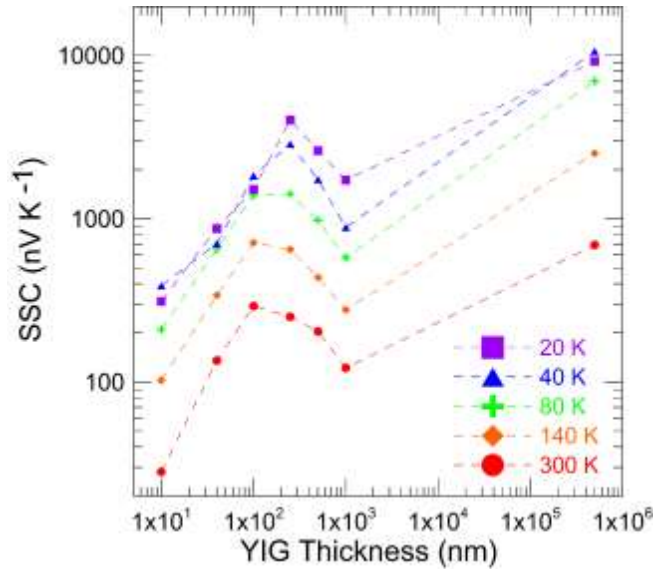

*Figure S4. Thickness dependence of the spin-Seebeck data in units of thermopower, using calculated values for the temperature gradient, shows slightly different temperature-dependent features than that of the data reported as a SSR, shown in Figure 2 of the main text.*

**Saturation magnetization study of samples with YIG thicknesses varying from 164 nm to 1000 nm.**

In this section we provide sample characterization data, in terms of saturation magnetization, obtained using a Vibrating Sample Magnetometer (VSM). Additional data from FMR linewidth measurements on films of thicknesses up to 250 nm can be found in Ref 14 and 15 of the main text. However, here we add a measurement showing the saturation magnetization of the 1 μm film to verify its quality. At room temperature, the magnetization of the 1 μm film appears to be less than that of the 250 and 164 nm films by roughly 7% (150 G), which is just outside the error bars of approximately 5% (100 G). This is reasonable, as 1 μm falls between thin film (< 250 nm) and bulk, so that the magnetization is starting to drop. The data sets show the same general trend versus temperature, with the same Curie temperature. Scanning Electron

Microscope (SEM) imaging of the 1 μm film confirms the measured thickness to be 1.04 μm +/- .04 μm.

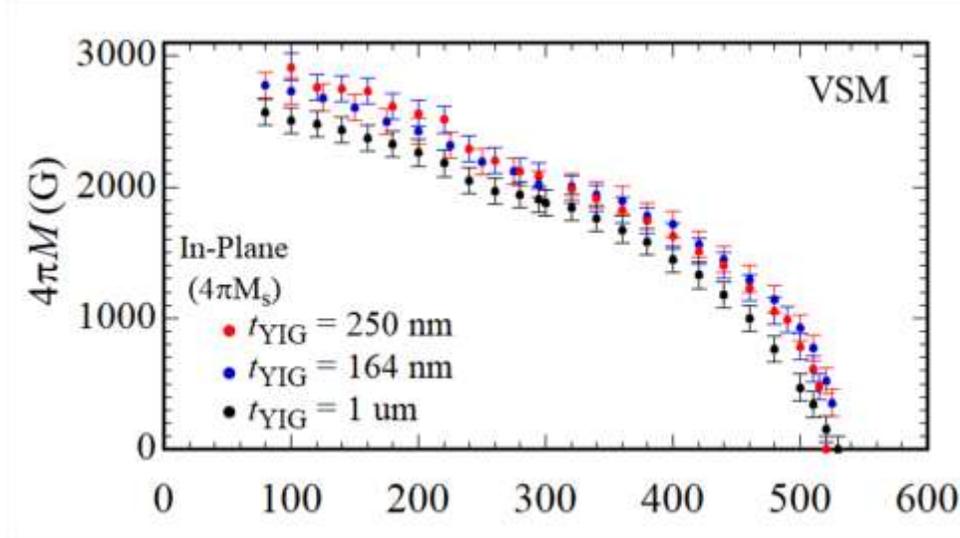

*Figure S5. Saturation magnetization as a function of temperature for the 1 μm YIG film, alongside measurements of the 250 nm YIG film. The 1 μm data show the same general trend and Curie temperature as the other samples with a slightly smaller magnitude, as can be reasonably expected from this thicker film.*

**Transport theory**

In this section, we introduce the equations describing the coupled magnon spin and heat transport in terms of a new set of length scales, and we offer an estimate for the latter. Looking for steady-state solutions and making use of Eqs. (3a) and (3b) of the main text, the spin and heat continuity equations (2a) and (2b) of the main text can be written as

$$\partial_x^2 \mu - \alpha_2 \partial_x^2 t = \frac{1}{\lambda_s^2}\mu - \frac{\beta_1}{\lambda_s^2}t, \tag{S1}$$

$$\partial_x^2 t - \alpha_1 \partial_x^2 \mu = t - \frac{\frac{\beta_1 \alpha_1}{\alpha_2}}{\lambda_s^2}\mu, \tag{S2}$$

where we have introduced $\beta_1 = g_{nT}/g_{n\mu}$, $\alpha_1 = \rho/\kappa$, and $\alpha_2 = \frac{\zeta}{\sigma}$. Here, both the chemical potential $\mu$ and the temperature difference $t = T_p - T_m$ are defined in units of $-\lambda_T \partial_x T_p > 0$, while spatial units have been normalized with respect to the magnon energy relaxation length $\lambda_T$. In the same spirit, for the boundary conditions (5a) and (5b) of the main text, one can rewrite:

$$\text{at } x = 0: \quad \alpha_2 + \alpha_2 \partial_x t - \partial_x \mu = -\frac{1}{\ell^*_{s,L}} \mu + \frac{\beta_2}{\ell^*_{s,L}} (\ell_L + t),$$

$$1 + \partial_x t - \alpha_1 \partial_x \mu = \frac{1}{\ell^*_{T,L}} (\ell_L + t) - \frac{\beta_2 \alpha_1/\alpha_2}{\ell^*_{s,L}} \mu, \tag{S3}$$

$$\text{at } x = d: \quad \alpha_2 + \alpha_2 \partial_x t - \partial_x \mu = 0,$$

$$1 + \partial_x t - \alpha_1 \partial_x \mu = \frac{1}{\ell^*_{T,R}} (\ell_R - t). \tag{S4}$$

Here, $\beta_2 = S_L/G_L$ and we have introduced the magnon thermal Kapitza length $\ell^*_{T,L(R)} = \kappa/K_{L(R)}$ at the Pt|YIG (YIG|GGG) interface, and the magnon spin Kapitza length $\ell^*_{s,L} = \sigma/G_L$ as the analogous length scale for spin transport.

Let us then write the spin current at the Pt|YIG interface as

$$j|_{x=0} = \frac{j_{s,q}}{\alpha_2 \ell^*_{s,L}} [\beta_2 (t|_{x=0} + \ell_L) + \mu|_{x=0}], \tag{S5}$$

where $j_{s,q} = -\zeta \partial_x T_p$ is the thermal spin current in the bulk.

For our analysis, we set the magnon temperature to $T_m \sim 150\,K$, which represents an average value of the temperature range that was spanned experimentally. We set $\ell_L/\ell_R \sim 1$ (Ref. [13] of the main text) and we take $\ell_R \sim 100\,nm$. Following the transport theory of Ref. [6] of the main text, we set $\alpha_{1,2} \sim \beta_{1,2} \sim 1$ and $\ell^*_{s,L} \sim 1\,\mu m$. Without further insights, we assume $\ell_L/\ell_R \sim \ell^*_{T,L}/\ell^*_{T,R} \sim 1$. We take the value of the spin-diffusion length, known experimentally from Ref. [3] of the main text, i.e., $\lambda_s \sim 10~\mu$m, and set $\lambda_s \sim 40~\lambda_T$, in agreement with the hierarchy $\lambda_T \ll \lambda_s$.

We remark that a fully diffusive approach to the coupled spin and heat transport may not capture quantitatively the salient transport features emerging on length scales that are shorter than the magnon mean free path. The latter has been estimated to exceed the micron scale at low temperatures (Ref. [17] of the main text). In order to gain a better qualitative understanding of the competing transport processes, we identify two distinct transport regimes as functions of the YIG thickness, namely $d \sim \lambda_T$ and $d \sim \lambda_s$. Our treatment relies on the energy-relaxation length scale being much shorter than the magnon spin-diffusion length, i.e., $\lambda_T \ll \lambda_s$. For thicknesses $d \sim \lambda_T \ll \lambda_s$, let us decouple heat from spin transport (i.e., by setting $\mu \sim 0$), so that the heat continuity equation simplifies to

$$\nabla^2 t = t. \tag{S6}$$

The latter has to be solved consistently with the following boundary conditions:

$$\text{at} = 0: \quad 1 + \partial_x t = \frac{1}{\ell^*_{T,L}}(\ell_L + t), \tag{S7}$$

$$\text{at} = d: \quad 1 + \partial_x t = \frac{1}{\ell^*_{T,R}}(\ell_R - t). \tag{S8}$$

The solution to Eqs. (S6), (S7), and (S8) reads as

$$t|_{x=0} = \frac{\ell^*_{T,L}(\ell_R - \ell^*_{T,R}) - (\ell_L - \ell^*_{T,L})(\ell^*_{T,R}\cosh d + \sinh d)}{(\ell^*_{T,R} + \ell^*_{T,L})\cosh d + \sinh d\,(1 + \ell^*_{T,R}\ell^*_{T,L})}, \tag{S9}$$

and is directly proportional to the injected spin current (S5), i.e.,

$$j|_{x=0} = \frac{j_{s,q}}{\alpha_2 \ell_{S,L}^*} [\beta_2(t|_{x=0} + \ell_L)] \tag{S10}$$

Equation (S10) has been plotted for short thicknesses in Fig. 5 of the main text by taking $\ell_{T,R}^* \sim 50\ nm$ and $\ell_{T,L}^* \sim 200\ nm$, and, for the sake of clarity, neglecting the second term on the right-hand side, i.e., the constant contribution given by the phonon Kapitza length $\ell_L$. We want to stress that while the non-monotonic behavior might be observed for $\ell_{T,R}^*/\ell_{T,L}^* \sim 1$ as long as $\ell_{T,R}^* \ll \lambda_T$, the peak structure can be accentuated further by decreasing $\ell_{T,R}^*$, which physically corresponds to enhancing the magnon-induced energy outflow at the right interface.

For samples of thickness $d$ comparable with the spin-diffusion length $\lambda_s$, we instead shift our focus to the pure spin transport, by anchoring the magnon temperature everywhere to the phonon temperature. The spin diffusion equation and the corresponding boundary conditions then reduce to:

$$\partial_x^2 \mu - \alpha_2 \partial_x^2 t = \frac{1}{\lambda_s^2} \mu, \tag{S11}$$

$$\text{at } x = 0:\ \alpha_2 - \partial_x \mu = -\frac{1}{\ell_{S,L}^*} \mu, \tag{S12}$$

$$\text{at } x = d:\ \alpha_2 - \partial_x \mu = 0. \tag{S13}$$

where we have neglected for simplicity the phonon Kapitza resistance. From Eq. (S11), (S12), and (S13) one easily can derive the injected spin current $j|_{x=0} = \frac{j_{s,q}}{\alpha_2 \ell_{S,L}^*} \mu|_{x=0}$ as

$$j|_{x=0} = -j_{s,q} \frac{\lambda_s}{\ell_{S,L}^*} \frac{1-\cosh(\frac{d}{\lambda_s})}{\frac{\lambda_s}{\ell_{S,L}^*}\cosh(\frac{d}{\lambda_s})+\sinh(\frac{d}{\lambda_s})}. \tag{S14}$$

The thickness dependence of Eq. (S14) is plotted in Fig. 4 of the main text.